\documentclass[preprint,12pt,authoryear]{elsarticle}

\usepackage{amssymb}

\usepackage{graphicx}
\journal{Planetary and Space Science}

\begin{document}

\begin{frontmatter}



\title{Spectra and physical properties of Taurid meteoroids}
\author[label1]{Pavol Matlovi\v{c}}\ead{matlovic@fmph.uniba.sk}
\author[label1]{Juraj T\'{o}th}
\author[label2]{Regina Rudawska}
\author[label1]{Leonard Korno\v{s}}
\address[label1]{Faculty of Mathematics, Physics and Informatics,
  Comenius University, Bratislava, Slovakia}
\address[label2]{ESA European Space Research and Technology Centre, Noordwijk, Netherlands}


\begin{abstract}
Taurids are an extensive stream of particles produced by comet 2P/Encke, which can be observed mainly in October and November as a series of meteor showers rich in bright fireballs. Several near-Earth asteroids have also been linked with the meteoroid complex, and recently the orbits of two carbonaceous meteorites were proposed to be related to the stream, raising interesting questions about the origin of the complex and the composition of 2P/Encke. Our aim is to investigate the nature and diversity of Taurid meteoroids by studying their spectral, orbital, and physical properties determined from video meteor observations. Here we analyze 33 Taurid meteor spectra captured during the predicted outburst in November 2015 by stations in Slovakia and Chile, including 14 multi-station observations for which the orbital elements, material strength parameters, dynamic pressures, and mineralogical densities were determined. It was found that while orbits of the 2015 Taurids show similarities with several associated asteroids, the obtained spectral and physical characteristics point towards cometary origin with highly heterogeneous content. Observed spectra exhibited large dispersion of iron content and significant Na intensity in all cases. The determined material strengths are typically cometary in the $K_B$ classification, while $P_E$ criterion is on average close to values characteristic for carbonaceous bodies. The studied meteoroids were found to break up under low dynamic pressures of 0.02 - 0.10 MPa, and were characterized by low mineralogical densities of 1.3 - 2.5 g\,cm\textsuperscript{-3}. The widest spectral classification of Taurid meteors to date is presented.
\end{abstract}


\end{frontmatter}

\section{Introduction}
\label{}

The Taurid complex is an extensive population of bodies associated with comet 2P/Encke and certainly one of the most interesting structures in the Solar System. It is the widest known meteoroid stream in the Solar System, and has been linked with several catastrophic incidents including the Palaeolithic extinctions (Napier, 2010) and the Tunguska event of 1908 (Kres\'{a}k, 1978). The low inclination of the orbit of the stream causes gravitational perturbations of inner solar system planets (Levison et al., 2006), resulting in the diffuse structure of the Taurid complex. The Earth encounters different parts of the stream annually from September to December, giving rise to several meteor showers, most significantly the Northern and the Southern Taurids peaking in late October and early November. Owing to the wide spread of the Taurid stream, the zenithal hourly rates of the resulting meteor showers usually do not exceed 5. However, the observed activity of the shower is known to be rich in bright fireballs, which induced discussions about the potential of Taurids in producing meteorites (Brown et al., 2013; Madiedo et al., 2014; Brown et al., 2016). While the orbits of most meteoroids are quite dispersed, it is still likely that the Taurid stream has a narrow and dense core consisting of particles concentrated near the orbit of the stream's parent object. It is often inferred that strong bombardment episodes have resulted at epochs when the material of the stream's core reached Earth intersection. 

Although comet 2P/Encke is generally considered as the parent object of the Taurids (Whipple, 1940), various small near-Earth objects (NEOs) and recently even two instrumentally observed carbonaceous meteorite falls have been linked with the orbit of the stream (Haack et al., 2011; Haack et al., 2012). 2P/Encke is a short-period comet moving on an orbit dynamically decoupled from Jupiter with an orbital period of 3.3 years. The peculiar orbit of the 2P/Encke raises interesting questions relating its origin and the origin of the Taurid complex. Today, the most supported hypothesis claims that comet 2P/Encke and several other bodies including large number of meteoroids were formed by a fragmentation of an earlier giant comet. The theory was first suggested by Whipple (1940), and later elaborated by Asher et al. (1993) who argued that the Taurid meteoroid streams, 2P/Encke, and the associated Apollo asteroids were all formed by major comet fragmentation 20 to 30 ky ago. Napier (2010) suggested that the debris of this fragmentation event could have caused the Palaeolithic extinctions followed by the return to ice age conditions.

The broad structure of the Taurid complex was studied by several authors with over 100 of NEO candidates associated with the Taurid complex. Porub\v{c}an et al. (2006) identified 15 different sub-streams of the complex and by applying stricter criteria for generic relations found associations with 9 NEOs. The presence of these bodies in the Taurid complex implies the possibility that part of the Taurid meteoroid population might be produced as the decay or impact products of the associated asteroids. The connection to carbonaceous chondrites represented by Maribo and Sutter's Mill meteorites also needs to be further investigated. We expect significantly different spectral and structural properties of such bodies in comparison to fragile cometary meteoroids originating in 2P/Encke. So far, the efforts in finding traces of the common origin of the largest Taurid complex bodies gave rather sceptical results. Popescu et al. (2014) studied the spectral properties of the largest asteroids associated with the Taurid complex, but found no evidence supporting mutual relation. Similarly, the spectroscopic and photometric measurements of Tubiana et al. (2015) found no apparent link between comet 2P/Encke, the Taurid complex NEOs, and the Maribo and Sutter's Mill meteorites.											

The Taurid meteor shower occasionally exhibits enhanced activity due to a swarm of meteoroids being ejected by the 7:2 resonance with Jupiter (Asher, 1991). The outburst of meteors observed in 2015 was anticipated by the model of Asher \& Izumi (1998), which previously predicted enhanced activities of Taurids in 1998, 2005 and 2008. Some features of the Taurid meteor shower activity suggest that the Taurid swarm exist only in the southern branch (Southern Taurids) and not in the northern branch (Shiba, 2016).									

The two main branches of the Taurid complex, the Northern and the Southern Taurids are well observed meteor showers with established orbital characteristics, which clearly trace them to comet 2P/Encke. The majority of Taurid shower analyses have been focused on orbital properties of the stream, with only several Taurid meteor spectra observed before the outburst of 2015 (e.g. Srirama Rao \& Ramesh, 1965; Madiedo et al., 2014). Borovi\v{c}ka et al. (2005) analyzed six Taurid emission spectra observed by low-resolution video spectrograph, identifying three Na-enhanced meteoroids, while another three Taurids have been classified as normal type. Using the same instrumentation, Voj\'{a}\v{c}ek et al. (2015) observed three more normal type Taurids including one meteoroid with lower iron content. The determined spectral characteristics of normal to enhanced sodium content and lower iron content would suggest cometary parent body. Asteroidal meteoroids are expected to be depleted in volatile sodium by the processes of space weathering (e.g. Borovi\v{c}ka et al., 2005; Trigo-Rodr\'{i}guez \& Llorca, 2007). Determined spectral properties give us valuable input into the studies of the origin of meteoroids; however, for doubtless differentiation between cometary and asteroidal particles, precise orbital and ideally physical characteristics must be obtained.

Besides the two major Taurid meteor showers, there are several smaller meteor streams associated with the Taurid complex. Most notably, this includes the Piscids, Arietids, chi Orionids, and the daytime showers of beta Taurids and zeta Perseids, encountered by the Earth in June and July. The beta Taurids and zeta Perseids have been shown to be the daytime twin branches to the Southern and the Northern Taurids respectively. The possibility of Taurid sub-streams being produced by a different parent object was discussed by Babadzhanov (2001), who found shower associations to each of the Taurid complex asteroids and interpreted them as the evidence for the cometary origin of these asteroids. Recently, Olech et al. (2016) found very close resemblance between the orbits of two 2015 Southern Taurid fireballs and the orbits of 2005 UR and 2005 TF50 asteroids. Precise observations of Taurid fireballs observed during the 2015 outburst were examined by Spurn\'{y} et al. (2016), who emphasized the orbital resemblance of the 2015 Taurids to the orbits of asteroids 2015 TX24 and 2005 UR. They argue that the outburst may have been caused by Taurid filaments associated with these asteroids.

All of the previous associations between different sub-streams of the Taurids and the Taurid complex asteroids were based on the orbital similarities. Certainly, studying the spectral and physical properties of the Taurid meteoroids could extend our understanding of the origin and evolution of the Taurid complex and its individual meteoroid streams.

\section{Observations and data reduction}
\label{}

The outburst of Taurid meteors in 2015 was observed globally by numerous meteor networks. Here we present the detailed analysis of 33 Taurid meteor spectra captured by the spectral All-sky Meteor Orbits System (AMOS-Spec) (Rudawska et al., 2016). Particular focus will be placed on 16 of these spectra, which were observed in Modra Observatory, Slovakia. Multi-station observations for 14 of these meteors were provided by four AMOS stations comprising the Slovak Video Meteor Network, which carries out routine meteor observations on every clear night (T\'{o}th et al., 2011; T\'{o}th at al., 2015), supplemented by individual observations provided by Pavel Spurn\'{y} (European Fireball Network) and Jakub Koukal (Central European MEteor NeTwork). 						

Additional 17 single-station Taurid spectra were observed by AMOS system in Chile during the testing for two new, now already established, southern AMOS stations. The system made observations during the activity peak between November 5 and November 10 at San Pedro de Atacama in the Atacama Desert. The observational conditions here enable much higher efficiency in capturing meteor events. These single-station meteors were identified as members of the Taurid stream based on their estimated radiant position and geocentric velocity.

The AMOS-Spec is a semi-automatic remotely controlled video system for the detection of meteor spectra. The main display components consist of 30mm f/3.5 fish-eye lens, image intensifier (Mullard XX1332), projection lens (Opticon 1.4/19 mm), and digital camera (Imaging Source DMK 51AU02). This setup yields a 100$^{\circ}$ circular field of view with a resolution of 1600 x 1200 pixels and frame rate of 12/s. The incoming light is diffracted by a holographic 1000 grooves/mm grating placed above the lens. The spectral resolution of the system varies due to the geometry of the all-sky lens with a mean value of 1.3 nm/px. A 500 grooves/mm grating was used for the observations in Chile, providing a mean spectral resolution of 2.5 nm/px. The system covers the whole visual spectrum range from app. 370 nm to 900 nm. The spectral response curve of the AMOS-Spec system (camera, image intensifier, and lens) is given in Figure 1. It was determined by measuring the known spectrum of Jupiter and is normalized to unity at 480 nm. The typical limiting magnitude of the system for meteors is approx. +4 mag., while only meteors brighter than approx. 0 mag. can be captured along with its spectrum. More details about the properties and capabilities of the AMOS systems can be found in T\'{o}th et al. (2015). The main disadvantage of the wide-field camera is the interference of the moonlight and bright Moon spectrum causing occasional difficulties in meteor detection. Fortunately, the Moon illumination percentage was descending from 45\% to 10\% during the activity peak of the 2015 Taurids (November 4 - November 8) and caused no significant problems in spectra analysis.

\begin{figure}
\centerline{\includegraphics[width=8cm,angle=0]{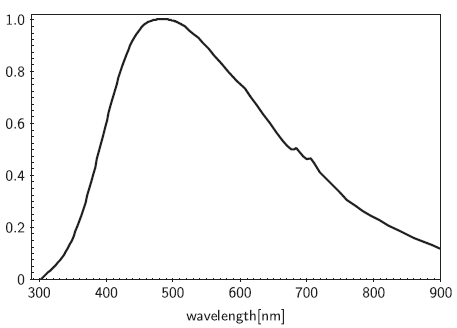}} \caption[f1]{The spectral sensitivity of the AMOS-Spec system. The spectral response curve was determined by measuring the spectrum of Jupiter, and is normalized to unity at 480 nm.} \label{response_curve}
\end{figure}

\subsection{Spectra reduction}
\label{}

The all-sky geometry of the lenses causes slight curvature of the meteor spectra captured near the edge of the FOV. For this reason, each spectrum was scanned manually on individual video frames using ImageJ\footnote{https://imagej.nih.gov/ij/} program. Before scanning the spectra, all of the analyzed spectral events were corrected for dark frame, flat-fielded, and had the star background image subtracted. We are particularly interested in the relative intensities of the Na I - 1, Mg I - 2, and Fe I - 15 multiplets, which form the basis of the spectral classification of meteors established by Borovi\v{c}ka et al. (2005). Each spectrum was fitted with a simple model accounting for all significant contributions to the observed spectral profile. The continuum level was fitted by Planck curve at given temperature. For most meteors, the continuum was well fitted by Planck curve at 4500 K, but in some cases, particularly for fainter meteors, this was lowered down. It should be noted that the fitting procedure does not serve as physical interpretation of the spectrum, but is only used for the reduction of contamination caused by the mixture of continuum radiation and weak unrecognizable lines, in order to achieve the best possible fit of each spectrum. The spectral lines (low temperature, high temperature, atmospheric, and wake lines) were fitted using the positions and intensities of the most important atomic lines in meteor spectra, as given in Borovi\v{c}ka et al. (2005). The simulated spectrum was fitted by instrumental Gaussian line profiles with appropriate full width of typically 5 nm at 1/e of the peak intensity. Furthermore, the most important N\textsubscript{2} bands of the first positive system present in the meteor spectra were fitted using the positions and intensities taken from Borovi\v{c}ka et al. (2005), and adjusted for our spectra. The Gaussian width of these bands was assumed to be 10 nm. The intensities of all contributions were adjusted in each spectrum to provide to best possible fit. The intensities of the main meteor multiplets (Mg I - 2, Na I - 1, and Fe I - 15) subtracted of the fitted continuous radiation and atmospheric emissions (most prominent lines of O, N, and N\textsubscript{2} bands) in the modeled spectrum were then measured and used in the spectral classification. The scale of each spectrum was determined using known spectral lines in the calibration spectrum and polynomial fitting of 2nd or higher degree. For most cases the lines of Mg I - 2 (518.2 nm), Na I - 1 (589.2 nm), O I - 1 (777.4 nm), and Fe I - 41 (438.4 nm) were used for the scaling of the spectrum, which provided optimal fit for the identification of other lines. Additionally, other prominent spectral lines were employed (e.g., Ca II - 1 at 393.4 nm, O I - 615.3 at nm, or N I - 3 at 744.2 nm.) in cases where other lines were missing in the FOV. The line identification, correction for spectral sensitivity, reduction of continuum radiation and atmospheric emission, and calculation of relative intensity ratios of studied spectral lines are all performed by our own developed Matlab code. Each spectrum was scanned and scaled on individual frames and the final profile was obtained by summing measured intensities on each frame. Only bright enough Taurid spectra with signal-to-noise ratio higher than 5 were used for our analysis.												

The low resolution of the video meteor spectra allows us only to identify limited number of meteoroid spectral lines compared to high dispersion photographic spectrographs used mainly in the past (e.g. Ceplecha, 1964; Nagasawa, 1978; Borovi\v{c}ka, 1993). It has been shown that meteor spectra consist of two spectral components (Borovi\v{c}ka, 1994). The majority of the observed meteoric spectral lines belong to the main, low-temperature ($\approx$ 4500 K) spectral component, which comprises mostly of the neutral lines of Na, Mg, Fe, Cr, Ca, and Mn. The high-temperature ($\approx$ 10000 K) spectral component is mostly significant in bright and fast meteors. It consists of atmospheric O and N lines, and meteoric high-excitation lines of Mg II, Si II, and Ca II. Only three meteoric species (Na I, Mg I, and Fe I) can be reliably examined in spectra captured by video spectrographs with resolution similar to our system. Even though other meteoric species are also often observed in meteor spectra (such as the lines of Ca I, Ca II, Si II, Cr I, or Mn I), only the multiplets of these three species have sufficient intensity in almost every meteor spectra to be accurately measured and compared. Nevertheless, the variations of relative intensities of these lines can reveal the different compositions of studied meteoroids. The full lists of the most important atomic lines identified in typical low resolution video meteor spectra can be found in Borovi\v{c}ka et al. (2005) or Voj\'{a}\v{c}ek et al. (2015).

\subsection{Photometry and astrometry}
\label{}

Previous measurements and orbital analyses of known meteor streams have demonstrated the astrometric precision of the AMOS system (T\'{o}th et al., 2015). The system achieves higher orbital accuracy compared to most all-sky video systems due to significantly higher pixel resolution. Multiple station observations of the analyzed spectral events were used to determine the heliocentric orbital parameters and geocentric trajectory parameters of the Taurid meteors. The AMOS systems comprising the SVMN yield a standard astrometric error of 0.03 - 0.05$^{\circ}$ that translates to the accuracy of tens to hundreds of meters for atmospheric meteor trajectory. The detailed all-sky reduction described by Ceplecha (1987) and Borovi\v{c}ka et al. (1995) was used. Each meteor was measured individually using UFOCapture software for meteor detection and UFOAnalyzer for astrometric data reduction (SonotaCo, 2009).   				

Magnitudes of meteors were determined by visual calibration based on comparison with bright stars, planets, and Moon phases in the FOV. These apparent magnitudes were then corrected to standard altitude of 100 km at the observation zenith, and for atmospheric extinction to obtain estimated absolute magnitudes. Using this method, magnitude of each studied meteor was determined from every available observing station, and resulting value (Table 5) was obtained as an average from individual station measurements. Absolute magnitudes of meteors captured in Chile (Table 2) were determined from single-station observations, which is why no error is given. Errors of these measurements from visual calibration are estimated to be $\pm 1$ magnitude for meteors of -5 magnitude and fainter, and $\pm 2$ magnitudes for brighter meteors. The errors of the estimated photometric masses are influenced besides the errors of absolute magnitudes and meteor velocities also by the effective meteoroid fragmentation model, which was not examined in our work. Therefore, the uncertainties of photometric masses is only estimated in given order of magnitude. Few exceptionally bright fireballs among the 2015 Taurids caused significant saturation, delivering higher uncertainty in the determination of the absolute magnitudes and consequently the photometric masses of the meteoroids.One of our multi-station meteors was observed under very low convergence angle, which translated into the low accuracy of the obtained orbital elements. These parameters were therefore omitted in our results (Table 5). In future, we plan to apply a new method based on detailed comparisons to the brightness of the Moon in various phases to gain higher accuracy of the estimated magnitudes. Furthermore, to gain the highest possible accuracy of the determined photometric masses, effective fragmentation model should be taken into account.

\section{Orbital classification}
\label{}

\begin{table}
\small\begin{center}\caption {Orbital properties of the multi-station Taurid meteor spectra observed during the outburst in November 2015. Each Taurid is designated with meteor number, meteor ID based on the observational date and time, shower assignment, corresponding Southworth-Hawkins criterion ($D_{SH}$) with respect to the assigned shower, geocentric velocity ($v_g$), semi-major axis ($a$), perihelion distance ($q$), eccentricity ($e$), inclination ($i$), argument of perihelion ($\omega$), longitude of the ascending node ($\Omega$), Tisserand's parameter with respect to Jupiter ($T_J$), and corresponding orbit type based on $T_J$ (AST for asteroidal orbits, JUP for Jupiter-family type orbits). Meteor no. 7 was omitted due to high uncertainty of the determined orbital elements, caused by the low convergence angle between the two observational stations. Meteor no. 15, and meteors no. 17 - 33 captured in Chile were only observed by one station, orbital properties are not available. } \vspace{0.5cm}
\resizebox{\textwidth}{!}{\begin{tabular}{ccccrrrrrrrrc}
\hline\\
&&&&&&&&&&\\[-12pt]
\multicolumn{1}{c}{No.}& %
\multicolumn{1}{c}{Meteor ID}& %
\multicolumn{1}{c}{Shower}& %
\multicolumn{1}{c}{$D_{SH}$}& %
\multicolumn{1}{c}{$v_g$}& %
\multicolumn{1}{c}{$a$} & %
\multicolumn{1}{c}{$q$} & %
\multicolumn{1}{c}{$e$} & %
\multicolumn{1}{c}{$i$} & %
\multicolumn{1}{c}{$\omega$}& %
\multicolumn{1}{c}{$\Omega$}& %
\multicolumn{1}{c}{$T_J$}& %
\multicolumn{1}{c}{type} \\
\hline\\
&&&&&&&&&\\[-12pt]
1 & M20151102\_020949 & STA & 0.08 & 28.74 & 1.87 & 0.316 & 0.831 & 5.5 & 120.19 & 39.10 & 3.45 & AST \\
  &  & & & $\pm$ 0.73 & 0.12 & 0.008 & 0.015 & 0.1 & 0.25 & & 0.19\\
2 &	M20151102\_024553 &	NTA & 0.14 & 27.62 & 2.04 & 0.365 & 0.821 & 3.3 & 293.69 & 219.16 & 3.27 & AST \\
  &  & & & $\pm$ 0.11 & 0.03 & 0.002 & 0.003 & 0.1 & 0.31 & & 0.04\\
3 &	M20151103\_212219 &	STA & 0.04 & 27.98 & 1.89 & 0.340 & 0.820 & 4.7 & 117.32 & 40.90 & 3.45 & AST \\
  &  & & & $\pm$ 0.35 & 0.06 & 0.004 & 0.008 & 0.1 & 0.03 & & 0.09\\
4 &	M20151103\_212454 &	STA & 0.10 & 29.27	& 1.90 & 0.305 & 0.839 & 6.6 & 121.13 & 40.90 & 3.40 & AST \\
  &  & & & $\pm$ 0.06 &	0.01 &	0.001 &	0.001 &	0.1 & 0.01 & & 0.02\\
5 &	M20151105\_205304 &	NTA & 0.05 & 30.12 &	2.35 &	0.314 &	0.866 &	1.8 &	298.08 & 222.95 & 2.89 & JUP \\
  &  & & & $\pm$ 0.32 &	0.09 &	0.003 &	0.007 &	0.1 &	0.03 & & 0.09\\
6 &	M20151105\_215813 &	STA & 0.03 & 28.98 &	2.39 &	0.354 &	0.852 &	4.3 &	113.45 & 42.92 & 2.89 &	JUP \\
  &  & & & $\pm$ 0.12 &	0.05 &	0.003 &	0.003 &	0.1 &	0.35 & & 0.04\\
8 &	M20151105\_220251 &	STA	& 0.03 & 28.34 &	2.17 &	0.356 &	0.836 &	3.7 &	114.04 & 42.92 & 3.11 &	AST \\
  &  & & & $\pm$ 0.17 &  0.07 &	0.005 &	0.004 &	0.7 &	0.72 & & 0.07\\
9 &	M20151105\_231200 &	STA	& 0.03 & 27.94 & 	2.02 &	0.356 &	0.824 &	4.9 &	114.78 & 42.98 & 3.28 &	AST \\
  &  & & & $\pm$ 0.38 &	0.08 &	0.004 &	0.009 &	0.1 &	0.06 & & 0.10\\ 
10 & M20151105\_235959 & STA & 0.17	& 26.59 &	2.55 &	0.438 &	0.828 &	0.8 &	103.60 & 42.92 & 2.83 &	JUP \\
  &  & & & $\pm$ 0.20 &	0.07 &	0.002 &	0.005 &	0.1 &	0.04 & & 0.05\\ 
11 & M20151108\_234416 & STA & 0.03	& 27.27 &	2.09 &	0.385 &	0.816 &	5.5 &	111.17 & 46.01 & 3.22 &	AST \\
  &  & & & $\pm$ 0.17 &	0.07 &	0.007 &	0.005 &	1.0 &	1.02 & & 0.08\\
12 & M20151109\_041942 & NTA & 0.05 & 29.30 &	2.21 &	0.330 &	0.851 &	1.0 &	296.74 & 226.32 & 3.04 & AST \\
  &  & & & $\pm$ 0.14 &	0.03 &	0.002 &	0.003 &	0.1 &	0.06 & & 0.04\\
13 & M20151111\_220940 & STA & 0.04	& 27.71 &	2.45 &	0.400 &	0.837 &	5.1 &	108.00 & 48.96 & 2.88 &	JUP \\
  &  & & & $\pm$ 0.15 &	0.05 &	0.002 &	0.004 &	0.1 &	0.01 & & 0.04\\
14 & M20151116\_193459 & STA & 0.10	& 25.51 &	2.08 &	0.441 &	0.788 &	4.9 &	104.98 & 53.88 & 3.28 &	AST \\
  &  & & & $\pm$ 0.04 &	0.01 &	0.000 &	0.001 &	0.1 &	0.01 & & 0.01\\
16 & M20151123\_012004 & NTA & 0.10 & 26.06	&   2.08 &	0.422 &	0.797 &	3.0 &	286.88 & 240.24 & 3.27 & AST \\
  &  & & & $\pm$ 0.03 &	0.01 &	0.001 &	0.001 &	0.1 &	0.20 & & 0.02\\

\hline
\end{tabular}}\label{Orbital}
\end{center}
\end{table}

The orbital properties of the 14 Taurid meteoroids observed by multiple stations are in Table 1. The Tisserand's parameter with respect to Jupiter can be used  to distinguish between the different orbit types (cometary, asteroidal) related to their source regions. It is defined as
\begin{equation}
T_J=\frac{a_J}{a}+2\sqrt{\Big(\frac{a}{a_J}\Big)(1-e^2)} \cos{i},
\end{equation}
where $a_J$ is the semi-major axis of Jupiter, and $a$, $e$, $i$ are the semi-major axis, eccentricity and inclination of the meteoroid orbit respectively. The Tisserand's parameter places our sample of meteoroids on the borderline between asteroidal orbits ($T_J >$ 3) and Jupiter-family type cometary orbits (2 $< T_J <$ 3), as would be expected for a stream originating in the short-period comet 2P/Encke ($T_J =$ 3.026). We identified four Jupiter-family type cometary orbits and ten asteroidal orbits among the studied meteoroids. Our sample includes particles from the anticipated outburst caused by the 7:2 resonance with Jupiter, along with the background activity of the Taurid stream. Based on the later discussed meteoroid properties (Table 5), we do not assume that found differences between meteoroids defined on asteroidal and Jupiter-family type orbits directly reflect different origin of these particles, but rather demonstrate the broad spatial structure of the stream, which is defined close to the borderline between these orbit classes. The determined Taurid orbits are displayed in Figure 2.			

\begin{figure}
\centerline{\includegraphics[width=8cm,angle=0]{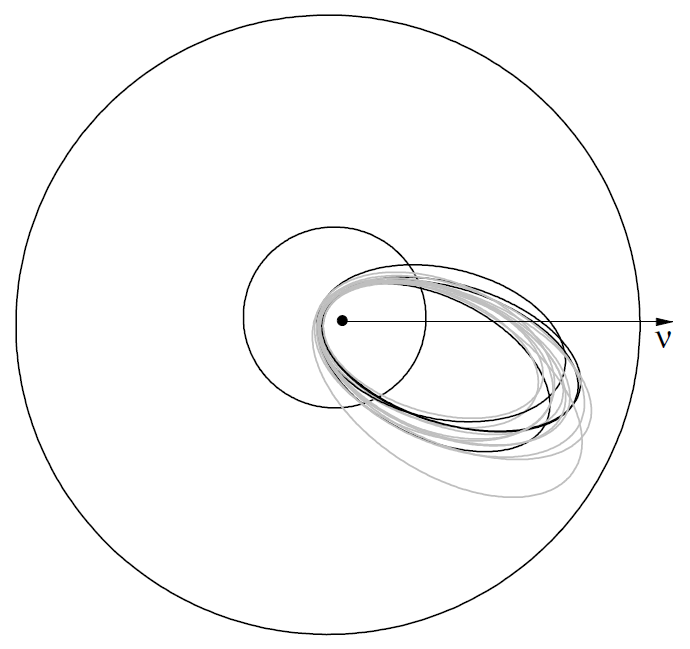}} \caption[f1]{Heliocentric orbits of 14 multi-station Taurid meteors based on orbital elements given in Table 1. Portrayed 4 Northern Taurid orbits (black), 10 Southern Taurids (grey), complemented with the orbit of Earth and Jupiter, and the direction towards the vernal equinox.} \label{Taurid_orbits}
\end{figure}

For further information about the orbital origin of our sample, we focused on using the Southworth-Hawkins $D_{SH}$ orbital similarity criterion (Southworth \& Hawkins, 1963) to look for associations of individual meteoroids with different filaments of the broad Taurid stream. Ten of the studied meteoroids were assigned to the southern branch of the stream, while four were identified as the Northern Taurids. Corresponding $D_{SH}$ values with respect to the assigned Taurid filaments (defined by orbital elements from Porub\v{c}an et al., 2006) are given in Table 1. Moreover, we also examined the orbital similarity between the studied meteoroids and the Taurid complex associated NEOs. Applying similarity threshold value of $D_{SH} \leq$ 0.10, we found possible relation to 10 asteroids: 2005 UR (associated with meteoroids no. 1, 4), 2015 TX24 (no. 1, 3, 4, 6, 8, 9), 2003 UV11 (no. 3, 6, 8, 9, 11, 13), 2007 UL12 (no. 3, 8, 9, 11, 13, 14), 2010 TU149 (no. 2, 6, 8, 10), 2003 WP21 (no. 14), 2010 VN139 (no. 14), 2004 TG10 (no. 5, 12), 2012 UR158 (no. 5, 12), and 2014 NK52 (no. 5, 12, 16). Compared orbital elements of the associated NEOs were taken from the Asteroid Orbital Elements Database currated by Ted Bowell and Bruce Koehn. The most associations in the southern branch of our sample was found with asteroids 2015 TX24, 2003 UV11, and 2007 UL12. Meteoroids from the northern branch showed most markable similarities with asteroid 2014 NK 52. These results support the measurements of Spurn\'{y} et al. (2016), who suggested that the 2015 outburst may have been caused by Taurid filaments associated with asteroids 2015 TX24 and 2005 UR. Two of the southern Taurid meteoroids (no. 1 and no. 4) originated on orbits closely resembling the orbit of asteroid 2005 UR ($D_{SH}$ = 0.06 and 0.07). The orbital similarity of 2005 UR with two bright Southern Taurid fireballs observed over Poland was also noted by Olech et al. (2016). The obtained orbital similarities with aforementioned asteroids are however still rather indefinite and could also be, given the broad structure of the Taurid complex, only incidental. In this work, we will focus on using the spectral and physical properties of Taurid meteoroids to look for features suggesting generic associations with asteroidal origin. 

\section{Spectral classification}
\label{}

As mentioned earlier, the low resolution of video spectrographs only allows us to study limited number of spectral lines in meteor spectra. However, the relative intensities of the three main meteoric species (Na, Mg, Fe) observed in the visual spectrum can reveal the rough composition of meteoroids. This is the basis of the spectral classification established by Borovi\v{c}ka et al. (2005), which distinguishes different types of bodies based on the intensities of spectral multiplets Na I - 1 (589.2 nm), Mg I - 2 (518.2 nm), and Fe I - 15 (526.0 - 545.0 nm). The modeled contributions of all recognized lines of the Fe I - 15 multiplet were summed. These multiplets were chosen as they can be well measured in most meteor spectra (see Section 2.1), and all lay in the region of high instrumental spectral sensitivity. Nevertheless, the measurement of these lines must be careful, because the Na multiplet overlaps with a sequence of N\textsubscript{2} bands, while the wake lines of Fe can influence the intensity of Mg line. All these effects were accounted for in the reduction process described in Section 2.1. As the spectral sensitivity of the AMOS-Spec is in this region similar to the sensitivity of the system used by Borovi\v{c}ka et al. (2005), we are able to apply the same classification also for our observations. It must be noted that the classification of Borovi\v{c}ka et al. (2005) was originally developed for fainter meteors in the magnitude range +3 to -1, corresponding to meteoroid sizes 1 - 10 mm. Our system observes meteor spectra of -1 to -10 magnitude, corresponding to meteoroid sizes of app. few mm to tens of cm. However, we expect that the same physical conditions fitted by the thermal equilibrium model applied for bright photographic meteors (Borovi\v{c}ka, 1993) as well as for fainter meteors (Borovi\v{c}ka et al., 2005), can be also assumed for the population observed by our system. The effect of self-absorption in brighter meteor spectra was examined in individual cases. The instrumental pixel saturation in spectra only occurred in individual meteors during bright flares and posed no problem for most meteors in the resolution provided by the 1000 grooves/mm grating. Individual frames with saturated meteor spectra were excluded from the summation for final spectral profiles and calculation of relative line intensities.	

\begin{figure}
\centerline{\includegraphics[width=12cm,angle=0]{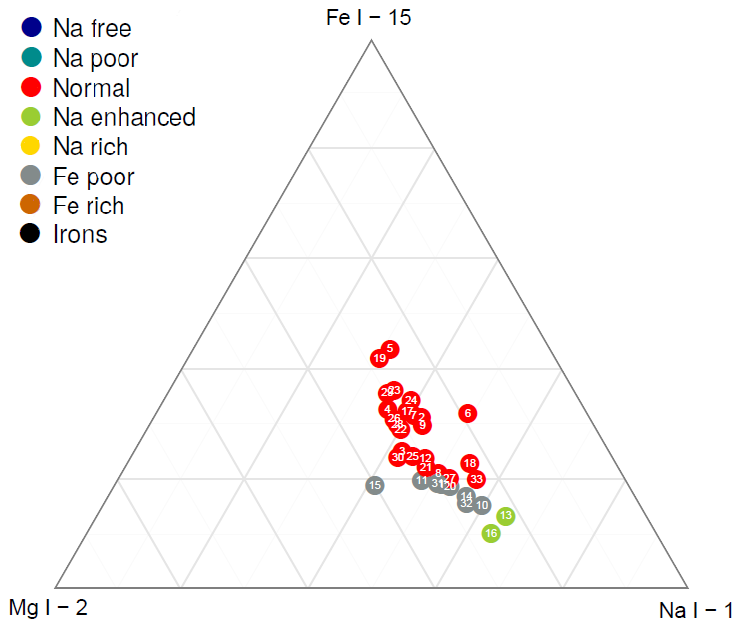}} \caption[f2]{The spectral classification of 33 Taurid meteors observed in Slovakia and Chile in November 2015. Diagram shows the measured relative intensities of multiplets Mg I - 2, Na I - 1, and Fe I - 15. Meteoroids are marked with numbers ordered according to the observational date. Meteors 1 - 16 were observed in Slovakia, while meteors 17 - 33 were captured in lower spectral resolution in Chile.} \label{spectral_classification}
\end{figure}

The spectral classification of Taurid meteoroids observed in Slovakia and Chile is on Figure 3. The majority (24) of Taurid meteoroids represent normal type spectral class with similar ratios of Na I, Mg I, and Fe I intensities. However, we observe apparent variations in Fe and Na content among the meteoroids in our sample. Particularly, the studied spectra revealed large dispersion of iron content in individual cases, ranging from Fe-poor bodies to almost chondritic Fe/Mg ratios (Figure 4). Normal type meteoroids are positioned in the middle of the ternary diagram, with content close to the ratios characteristic for chondritic composition. It was noted by Borovi\v{c}ka et al. (2005) and later confirmed by Voj\'{a}\v{c}ek et al. (2015) that the majority of observed meteoroids fall in the ternary diagram below the so called chondritic curve, which represents the expected range for chondritic composition, due to the lower content of iron. This is also the case for our sample of Taurid meteoroids, though two of the measured spectra showed quite significant Fe intensities (meteoroids no. 5 and 19). Furthermore, we also identified 8 Taurids depleted in iron, representing Fe-poor class, and two Na-enhanced Taurid meteoroids (no. 13 and 16). As can be clearly seen in Figure 3, all of the Taurids are positioned right from the center of the ternary diagram, exposing slight Na line intensity enhancement over Mg in all of the studied meteoroids. This supports the results of Borovi\v{c}ka et al. (2005), who found Na-enhancement in three of the 6 observed Taurid spectra.

\begin{figure}
\centerline{\includegraphics[width=9cm,angle=0]{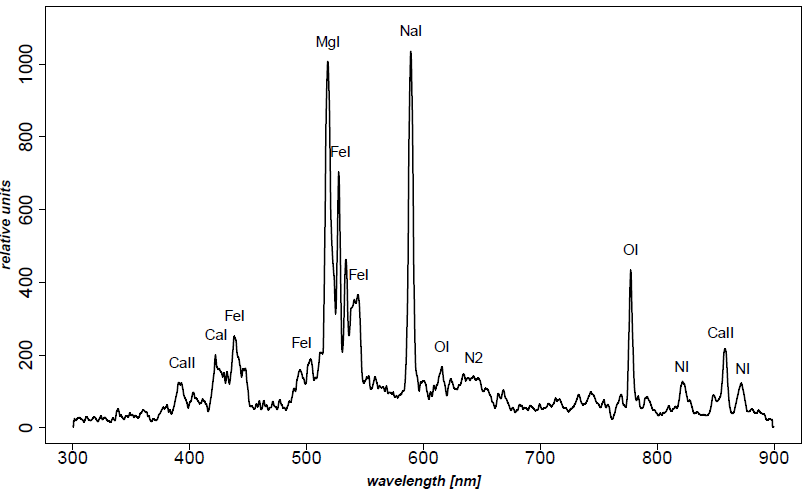}}
\centerline{\includegraphics[width=9cm,angle=0]{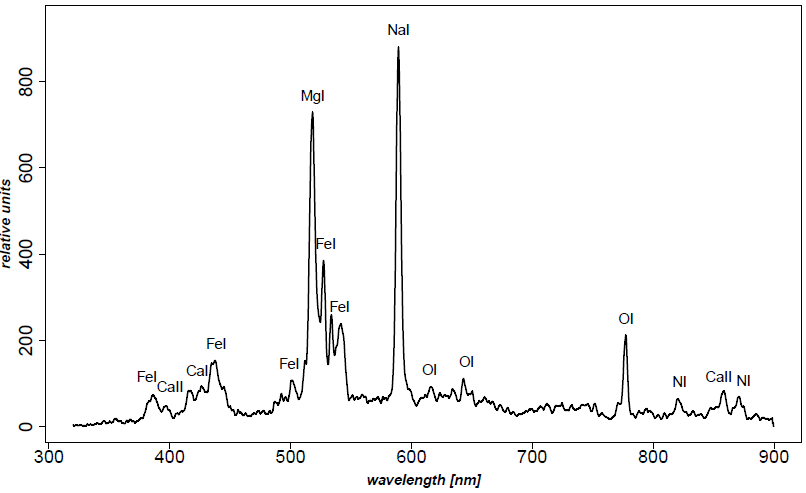}}
\centerline{\includegraphics[width=9cm,angle=0]{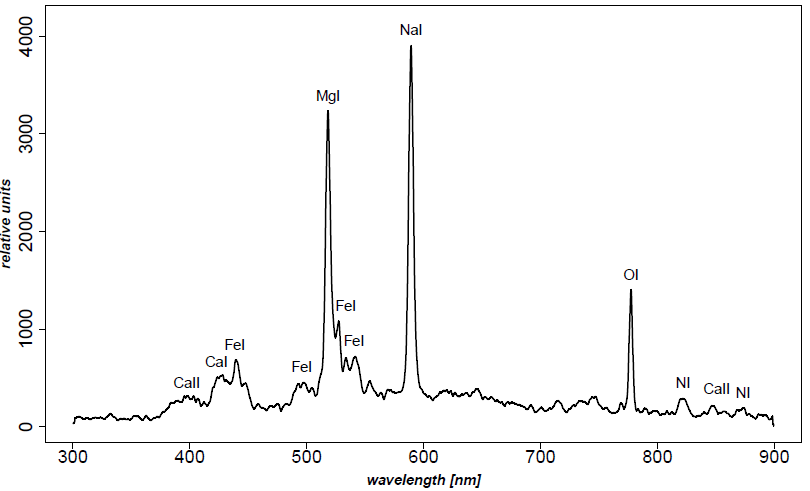}} \caption[f4]{Spectral profiles of Taurid meteoroids no. 5, 7, and 11. The spectra show similar features with significant dispersion of the Fe I - 15 multiplet intensity from the highest (no. 5) to the lowest (no. 11). Displayed spectral profiles are not corrected for the spectral sensitivity of the system (Fig. 1).} \label{iron_variation}
\end{figure}

The Na line intensity is certainly one of the most interesting features in meteor spectra, revealing the thermal history of meteoroids. The presence of sodium in meteor spectra can be used as a tracer of volatile phases associated with cometary origin. During the formation of the solar system, volatile elements were depleted from the inner protoplanetary disk due to intense solar radiation (Despois, 1992). We expect higher Na abundance in unprocessed bodies such as comets. Borovi\v{c}ka et al. (2005) pointed out three populations of Na-free meteoroids: iron bodies on asteroidal orbits; meteoroids with small perihelia ($q \leq$ 0.2 AU), where Na was lost by thermal desorption; and Na-free meteoroids on Halley type orbits, where Na loss was possibly caused by irradiation of cometary surfaces by cosmic rays in the Oort cloud. The depletion in Na for small meteoroids is observed as a function of the time exposure to solar radiation, which predicts that asteroidal particles will be Na-poor materials (Trigo-Rodr\'{i}guez \& Llorca, 2007). Even though all of our Taurid spectra have shown unambiguous sodium content, we observe apparent effect of Na depletion as a function of perihelion distance (Figure 5). The observed dependence of Na/Mg ratio on perihelion distance however can be partially influenced by the effect of saturation/self-absorption in bright (larger) meteors, while the measurement of line intensity in fainter spectra (typical for smaller particles) is a subject of higher uncertainty. The errors of determined Na/Mg and Fe/Mg intensity ratios from signal-to-noise ratio in individual observations are given in Table 5. As can be seen in Figure 5, the Na enhancement was preferred in smaller Taurid meteoroids, in which the Na/Mg ratio was determined with higher uncertainty. Furthermore, meteoroids with high surface-area/volume ratios are expected to suffer higher depletion of Na and other volatiles through solar radiation (Trigo-Rodr\'{i}guez \& Llorca, 2007).

\begin{figure}
\centerline{\includegraphics[width=8cm,angle=0]{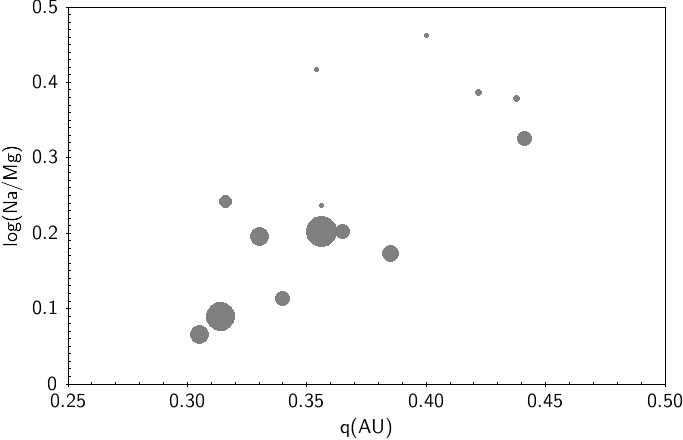}} \caption[f4]{The dependence of the Na/Mg line intensity ratio on perihelion distance. The varying sizes of the marks display photometric masses of individual meteoroids in logarithmic scale.} \label{Na_q_size}
\end{figure}

The increased values of Na/Mg ratios in Taurid spectra are not too surprising. It was pointed out that the Na/Mg line intensity ratio varies widely in cometary meteoroids (e.g. Borovi\v{c}ka et al., 2005; Rudawska et al., 2016), and is observed as a function of meteor velocity. Sodium is characterized by a low ionization potential, and the Na line is of low excitation (2.1 eV) compared to the Mg line (5.1 eV) (Borovi\v{c}ka et al., 2005). Consequently, we expect Na line intensity to be dominant at lowest velocities (temperatures) and lower at highest velocities (temperatures). This effect is only observed for meteor velocities below 40 km/s and increases for velocities below 20 km/s. Typical Taurid velocities of approx. 28 km/s would therefore slightly favor Na intensity over Mg, particularly for fainter meteors. Bright fireballs caused by more massive meteoroids can reach very high temperatures, at which both Na and Mg become fully ionized, as observed in Figure 5. 

Moreover, also the intensities of atmospheric emission lines of O I and N I, as well as the N\textsubscript{2} bands increase with the meteor velocity. As a result of the moderate velocities of the Taurid meteors, the atmospheric emission lines were clearly present in the spectra (Figure 4), but did not achieve high intensities, allowing us to readily subtract them from the spectral profiles before determining the intensities of the main meteoroid multiplets.	

The brightest multi-station Taurid in our sample was observed on November 5 at 23:21:00 UT. This fireball (marked no. 9 in Figure 3) of estimated -8.4 absolute magnitude exhibited bright flare, which enabled us to capture two orders of its spectrum. The 1st order and the 2nd order spectral profiles are on Figure 6. While the 1st order spectrum unveils more detailed features with emission lines of the high-temperature spectral component, the saturation in several frames caused by the flare restricts us from analyzing the relative intensities of the main spectral multiplets throughout the entire meteor flight. For this reason, we used the higher dispersion 2nd order spectrum to determine the spectral classification of this Taurid. Clearly, the relative intensities of Mg I - 2, Na I - 1, and Fe I - 15 differ in the two spectral profiles. We can observe the lines of Fe I - 15 blending into the Mg I - I radiation and increasing its intensity. This demonstrates the uncertainty of determining relative intensities in bright spectra of meteor flares.	

\begin{figure}
\centerline{\includegraphics[width=9cm,angle=0]{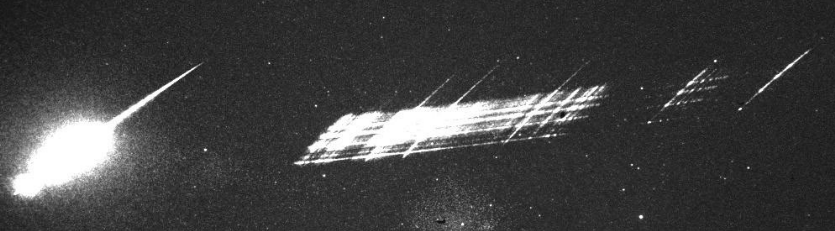}}
\centerline{\includegraphics[width=9cm,angle=0]{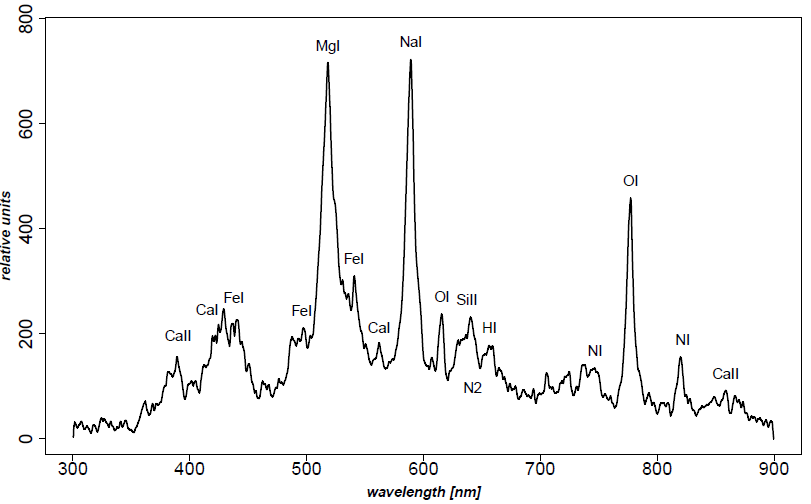}}
\centerline{\includegraphics[width=9cm,angle=0]{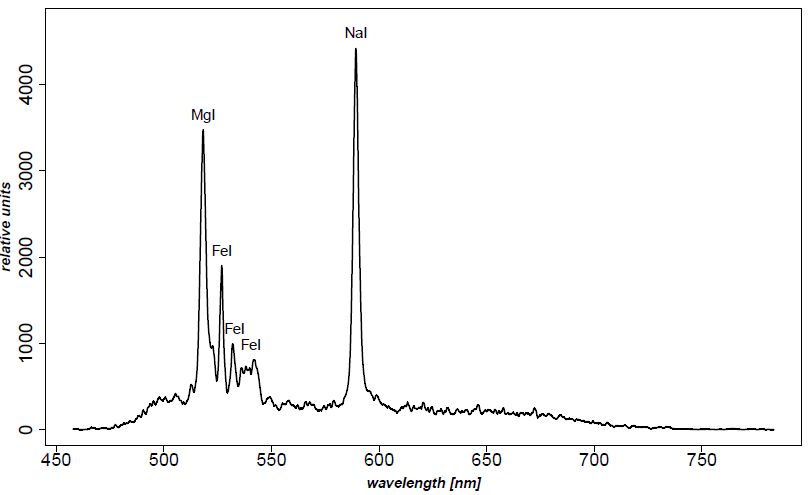}} \caption[f5]{Brightest multi-station Taurid meteor (no. 9) in our sample of estimated -8.4 magnitude captured on November 5 at 23:21:00 UT, and the spectral profiles of the first order spectrum (upper picture) and second order spectrum (lower picture) showing different relative intensities of the Mg I - I, Na I - 2, and Fe - 15. multiplets. Displayed spectral profiles are not corrected for the spectral sensitivity of the system (Fig. 1).} \label{no9}
\end{figure}

Example of a typical lower-resolution Taurid spectrum captured by a system in Chile using the 500 grooves/mm grating is on Figure 7. Even though we are not able to study fainter spectral features in these profiles, the intensities of the main spectral multiplets are well measurable. The blending of Mg I - 2 and Fe I - 15 multiplets is more apparent in low-resolution spectra (Figure 7), but it was resolved for the relative intensity measurement by fitting the resulting profile shape with intensities of all expected spectral lines in given area. The general results are in good agreement with the spectral classification of Taurids based on observations in Slovakia. Estimated absolute magnitudes of studied single-station Taurid meteors captured with spectra in San Pedro de Atacama Chile are in Table 2.

\begin{figure}
\centerline{\includegraphics[width=9cm,angle=0]{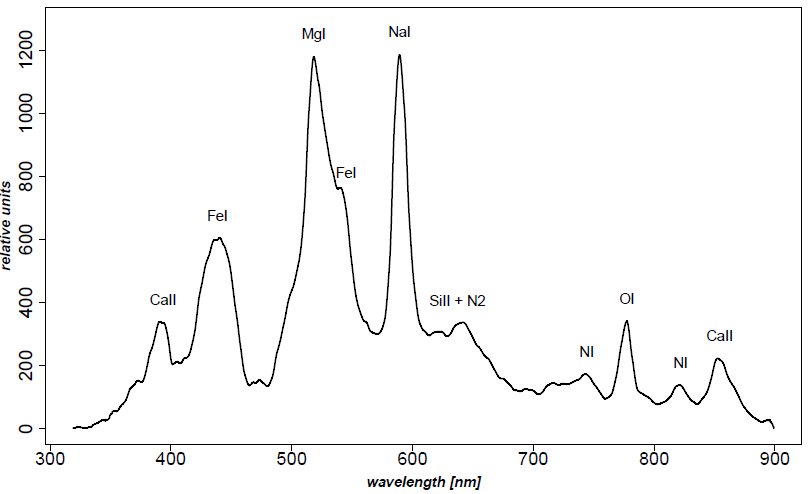}} \caption[f6]{Example of a low-resolution Taurid spectrum (no. 29) captured on November 8 at 07:00:43 UT in Chile. Displayed spectral profile is not corrected for the spectral sensitivity of the system (Fig. 1).} \label{no29}
\end{figure}

Generally, the spectral characteristics of the measured Taurid spectra, which are defined by mostly normal-type composition with Fe content below the chondritic values and increased Na content in most meteors point towards the cometary origin of all of the studied Taurid meteoroids. Nevertheless, we observe large variations of Fe and notable Na dispersion in Taurid spectra, suggesting Taurids are very heterogeneous population of meteoroids. Two of the Taurid spectra exhibited slightly enhanced Fe intensities (no. 5 and no. 19). The spectrum of meteor no. 19 was captured from a short, bright flare (estimated -9.4 absolute magnitude) and in lower spectral resolution, which brings higher uncertainty of measured relative intensities related to possible effects of saturation/self-absorption. The observed heterogeneity is not unexpected for a stream of such remarkable spatial dispersion. In addition to the original heterogeneity of the parent body of the stream, the diverse thermal evolution in different parts of the stream would result in the observed compositional variations. We did not find any spectral features, which would clearly indicate asteroidal origin in any of the meteoroids in our sample, though the relative intensities Na I, Mg I, and Fe I in several meteoroids were not too distant from the values expected for chondritic bodies. For more information, we focused on using the multi-station observations of 14 Taurid meteors to determine the physical characteristics of the studied meteoroids.

\begin{table}
\centering
\small\begin{center}\caption {Estimated absolute magnitudes ($Mag$) of single-station Taurid meteors captured with spectra in San Pedro de Atacama, Chile. Each meteor is designated with meteor number and meteor ID based on the observational date and time. Errors of absolute magnitudes are estimated to be $\pm 1$ magnitude for meteors of -5 magnitude and fainter, and $\pm 2$ magnitudes for brighter meteors.} \vspace{0.5cm}
{\begin{tabular}{lcccccccccc}
\hline\\
&&&\\[-12pt]
\multicolumn{1}{l}{No.}& %
\multicolumn{1}{c}{Meteor ID}& %
\multicolumn{1}{c}{$Mag$}&  \\
\hline\\
&&\\[-12pt]
17 & M20151106\_042700 &	-5.5 \\
18 & M20151106\_052449 &	-4.5 \\
19 & M20151106\_062854 &	-9.4 \\
20 & M20151106\_063233 &	-3.5 \\
21 & M20151106\_070343 &	-8.6 \\
22 & M20151107\_031208 &	-3.9 \\
23 & M20151107\_061350 &	-7.7 \\
24 & M20151107\_063741 &	-7.8 \\
25 & M20151107\_072801 &	-5.5 \\
26 & M20151107\_073808 &	-5.2 \\
27 & M20151108\_020459 &	-2.2 \\
28 & M20151108\_025443 &	-11.6 \\
29 & M20151108\_070043 &	-10.4 \\
30 & M20151108\_071904 &	-2.8 \\
31 & M20151108\_083951 &	-4.9 \\
32 & M20151110\_004036 &	-6.8 \\
33 & M20151110\_054715 &	-4.1 \\
\hline
\end{tabular}}\label{Laboratory}
\end{center}
\end{table}

\section{Physical properties of Taurid meteoroids}
\label{}

The primary focus of this section will be placed on the material strengths of Taurid meteoroids, which could indicate additional asteroidal source of meteoroids in the stream, and also tell us about the potential of this stream in producing meteorites.

The first physical parameter we need to determine is the meteoroid mass, which is also later used in the calculation of other physical parameters. The activity of the Taurid meteor shower is characteristically rich in bright fireballs indicating larger particles than what we observe in most cometary showers (Wetherill, 1974). It was suggested that Taurids contain very large meteoroids at least in the hundreds of kilograms range, which is more than an order of magnitude larger than other showers (Brown et al., 2013; Ceplecha et al., 1998). Our sample included Taurid meteors ranging from -1.3 to -8.4 magnitude, only confirming these characteristics. The meteoroid masses were determined photometrically, i.e. by integrating along the light curves of meteors. The standard luminosity equation (e.g. Bronshten, 1983) can be rewritten in the form:
\begin{equation}
m_\infty=2\int_{t_B}^{t_E} \frac{Idt}{\tau_l v^2}.
\end{equation}
Here, $m_\infty$ is the original mass of the meteoroid, $v$ is the meteor velocity, and $I$ is the light intensity of the meteoroid, integrated in the time interval from the beginning of the light curve ($t_B$) to the end ($t_E$). The luminosity efficiency factor, $\tau_l$, represents the fraction of the meteoroid's kinetic energy, which has been converted into luminosity in the visual range.  For the calculation of $\tau_l$, we applied the approach described by Hill et al. (2005). The method uses a relation between the luminous efficiency factor and universal excitation coefficient $\zeta$ defined by Jones \& Halliday (2001):
\begin{equation}
\tau_l=2\Big(\frac{\epsilon}{\mu}\Big)\frac{\zeta}{v^2},
\end{equation}
where the values of $\zeta$ are split into several regimes based on the meteor velocity and the ratio of the mean excitation energy to the atomic mass ($\epsilon/\mu$) is computed for meteoric atoms and ions producing the most radiation (e.g. Ceplecha et al., 1998), assuming the majority of the observed radiation comes from the meteoroid atoms (Bronshten, 1983). The estimated masses of the meteoroids in our sample range from 1 to 1500 grams (Table 5). 

\subsection{Material strength}
\label{}

Meteor observations can be used to characterize the material strength of meteoroids by determining empirical parameters $K_B$ and $P_E$, which are functions of the beginning and the terminal height of the meteor luminous path. Fragile particles tend to crumble under lower pressures resulting in higher beginning height of the observed luminous trajectory. On the contrary, only stronger materials can withstand the pressure and reach lower parts of our atmosphere, and only the most robust particles will eventually fall on the Earth's surface. It is observed that depending on the entry velocity and meteoroid mass, the ablation begins at higher heights for cometary bodies than asteroidal ones (Ceplecha \& McCrosky, 1976; Koten et al., 2004).

The $K_B$ and $P_E$ parameters defined by Ceplecha (1968), and Ceplecha \& McCrosky (1976), and later revised by Ceplecha (1988) can be expressed as:
\begin{equation}
K_B=\log{\rho_B} + 2.5 \log{v_\infty} - 0.5 \log{z_R},
\end{equation}
\begin{equation}
P_E=\log{\rho_E} - 0.42 \log{m_\infty} + 1.49 \log{v_\infty} - 1.29 \log{\cos{z_R}}.
\end{equation}
Here, $\rho_B$ and $\rho_E$ are the atmospheric densities at beginning and end height of the luminous trajectory, $v_\infty$ is the pre-atmospheric meteor velocity, $z_r$ is the zenith distance of the radiant, and  $m_\infty$ is the original photometric mass calculated using luminous efficiency factor in accordance with Ceplecha \& McCrosky (1976). The values of atmospheric densities at different heights were obtained using the MSIS-E-90 Atmosphere Model of Hedin (1991). These parameters form a basis for the empirical classification of meteoroid material strength (Table 3). The $K_B$ classification differentiates between fragile cometary bodies (group D) characteristic e.g. for Draconid meteors, typical cometary bodies (group C), dense cometary bodies (group B), carbonaceous-type bodies (group A), and strong meteoroids of asteroidal origin (group ast). Similarly, the classification based on the $P_E$ parameter distinguishes between cometary bodies (group III), carbonaceous-type bodies (group II), and asteroidal bodies (group I). 

\begin{table}
\small\begin{center}\caption {Meteoroid material types based on material strength parameters $K_B$ and $P_E$, as defined by Ceplecha (1988). The differences between groups C3 and C2, IIIAi and IIIA are based on orbital properties (for further information, see Ceplecha, 1988)} \vspace{0.5cm}
\resizebox{\textwidth}{!}{\begin{tabular}{ccccc}
\hline\\
&&&&\\[-12pt]
\multicolumn{1}{c}{Material type}& %
\multicolumn{1}{c}{$K_B$ group}& %
\multicolumn{1}{c}{$K_B$}& %
\multicolumn{1}{c}{$P_E$ group}& %
\multicolumn{1}{c}{$P_E$} \\
\hline\\
&&&\\[-12pt]
Fragile cometary & D & $K_B <$ 6.6 & IIIB & $P_E \leq$ -5.70 \\ [2pt]
\begin{tabular}{@{}c@{}}Regular cometary \\ long-period\end{tabular}  & C3 & 6.6 $\leq K_B <$ 7.1 & - & - \\
\begin{tabular}{@{}c@{}}Regular cometary \\ long-period\end{tabular} & C2 & 6.6 $\leq K_B <$ 7.1 & IIIAi & -5.70 $< P_E \leq$ -5.25 \\
\begin{tabular}{@{}c@{}}Regular cometary \\ short-period\end{tabular} & C1 & 6.6 $\leq K_B <$ 7.1 & IIIA & -5.70 $< P_E \leq$ -5.25 \\
Dense cometary & B & 7.1 $\leq K_B <$ 7.3 & - & - \\ [2pt]
Carbonaceous chondrites & A & 7.3 $\leq K_B <$ 8.0 & II & -5.25 $< P_E \leq$ -4.60 \\ [2pt]
\begin{tabular}{@{}c@{}}Ordinary chondrites \\ Asteroids\end{tabular} & ast & 8 $\leq K_B$ & I & -4.60 $< P_E$ \\
\hline
\end{tabular}}\label{KBPE}
\end{center}
\end{table}

\begin{figure}
\centerline{\includegraphics[width=13.5cm,angle=0]{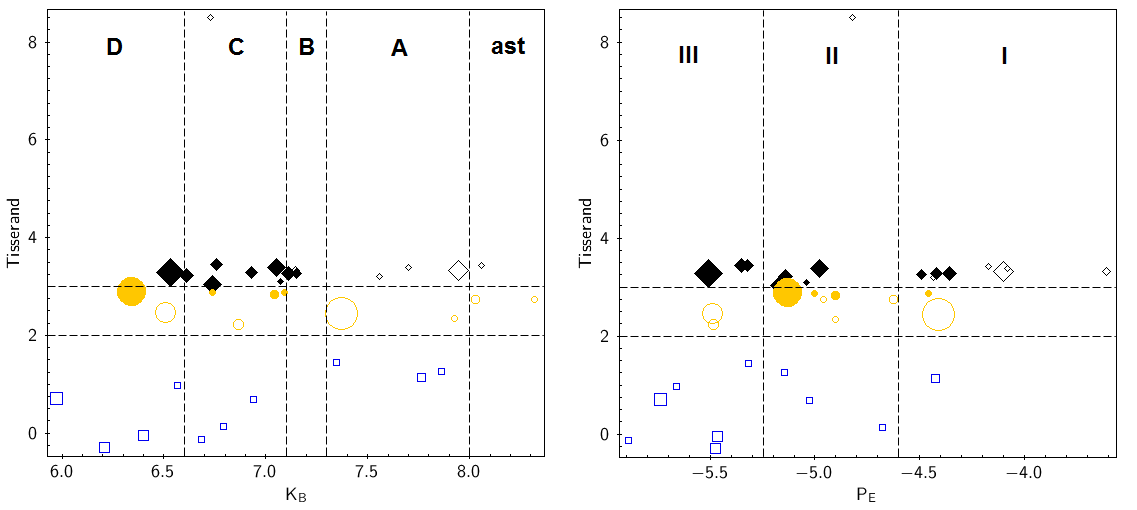}} \caption[f7]{Classification of meteoroid material strength based on parameters $K_B$ and $P_E$ compared to the orbital classification of meteoroids based on Tisserand’s parameter. Taurid meteoroids (filled marks) are compared to diverse population of 22 meteoroids taken from Rudawska et al. (2016). Blue marks represent meteoroids on Halley-type cometary orbits, yellow marks designate Jupiter-family cometary orbits, and black represent asteroidal orbits. The varying sizes of the marks display photometric masses of individual meteoroids in logarithmic scale.} \label{KBPEALL}
\end{figure}

We clearly observe a large scatter of material strengths among Taurid meteoroids in our sample (Table 5). Following the $K_B$ classification (Figure 8), the majority of our Taurids fall into the C group characteristic for standard cometary bodies, with only two meteoroids appearing to be more fragile (D group) and two meteoroids of slightly higher strength (B group). It should be noted that the $K_B$ parameter (Eq. 4) assumes that beginning height does not depend on meteoroid mass (meteor magnitude). While this is true for a narrow range of meteor magnitudes, Koten et al. (2004) have shown that the beginning height increases with mass for most meteor showers. The classification of the two brightest Taurids in our sample as type D is probably consequence of this fact. The obtained $K_B$ values are nevertheless what we would expect from cometary bodies originating in 2P/Encke. However, we see significantly higher dispersion of Taurid material strengths in the $P_E$ classification (Figure 8). Here, only three meteoroids have $P_E$ values characteristic for cometary bodies, while the majority is defined as group II, similar to carbonaceous bodies, and four Taurids even exposed end-height strengths typical for rocky asteroidal bodies (group I). Similarly, Brown et al. (2013) found that most Taurids appear to be type II or type III fireballs. Great variations of material strength in Taurid meteoroids including very strong objects among fireballs were previously indicated by Borovi\v{c}ka (2006). It seems that while the beginning heights of Taurids are typically cometary, the end heights are usually lower and more dispersed (Figure 9). Ceplecha (1988) noted that parent bodies of group II and group A meteoroids are perhaps partly asteroids and partly comets (Wetherill \& Revelle, 1981). However, the discovered Taurid features more likely suggest heterogeneous structure of its meteoroids, which are of cometary nature but contain solid, possibly carbonaceous inclusions. The diversity of material types observed on Taurid orbits could be explained by the inhomogeneous interior of the parent comet (2P/Encke or earlier larger body). The existence of solid inclusions, which relate to CI and CM carbonaceous chondrites has also been indicated in a small population of Perseid meteoroids with high mineralogical densities (Babadzhanov \& Kokhirova, 2009), and confirmed in the material of Leonid meteoroids (Borovi\v{c}ka, 2006). Surprisingly low material strength is observed in meteoroid no. 5, which exhibited the highest content of iron in our sample. While this is the second brightest Taurid (estimated -8.1 magnitude), no signs of significant effects of saturation/self-absorption, which could cause over-estimation of iron intensity, were observed in the spectrum. There is no clear evidence supporting asteroidal origin in any of the Taurid meteoroids, nevertheless, the connection to carbonaceous chondrites remains unrefuted. Furthermore, the diversity of material strengths among Taurids could also relate to the thermal history of these meteoroids. It is usually observed that meteoroids depleted in sodium are stronger than other particles (Borovi\v{c}ka et al., 2005). However, according to the $P_E$ classification, the two Na-enhanced meteoroids in our sample (no. 13 and no. 16) were both identified as group I meteoroids (Table 5). Also the $K_B$ classification of Taurid meteoroids suggests that particles with slightly enhanced Na/Mg ratios were generally characterized by higher material strength. One explanation of this effect could be the exposure to stronger solar radiation in lower perihelia orbits, which may have caused slight structural disruptions, resulting in lower material strengths observed in $K_B$ classification. The dependence of material strength parameter $K_B$ on sodium content (Figure 10) is apparent in the majority of the sample, but was not manifested in a branch of four meteoroids, suggesting this effect has only partial influence on the observed material strength. The observed effect could be however also influenced by the depletion of Na/Mg in brighter meteors, possibly affected by saturation/self-absorption, as was mentioned earlier. The overall physical properties displayed in Table 5 indicate four meteoroids of group I, also characterized with higher mineralogical densities, as the best possible candidates for non-cometary origin, however, the asteroidal source cannot be confirmed in any of the studied meteoroids. 

\begin{figure}
\centerline{\includegraphics[width=8cm,angle=0]{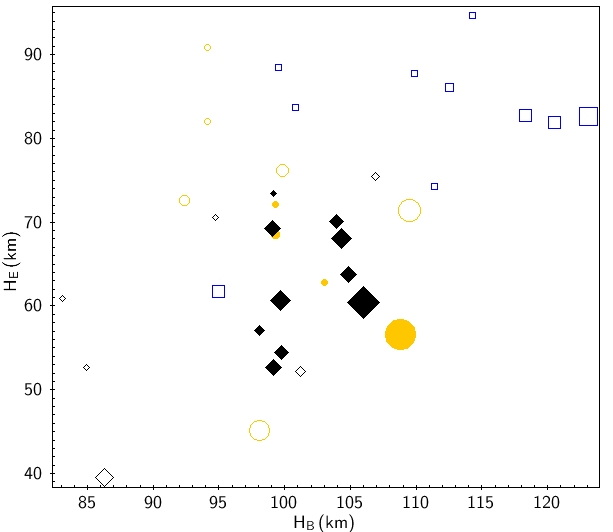}} \caption[f8]{Comparison of the beginning and terminal heights of meteors. Taurid meteoroids (filled marks) are compared to diverse population of 22 meteoroids taken from Rudawska et al. (2016). The marking of meteoroids is the same as in Figure 8.} \label{Hb_He_logmp}
\end{figure}

\begin{figure}
\centerline{\includegraphics[width=8cm,angle=0]{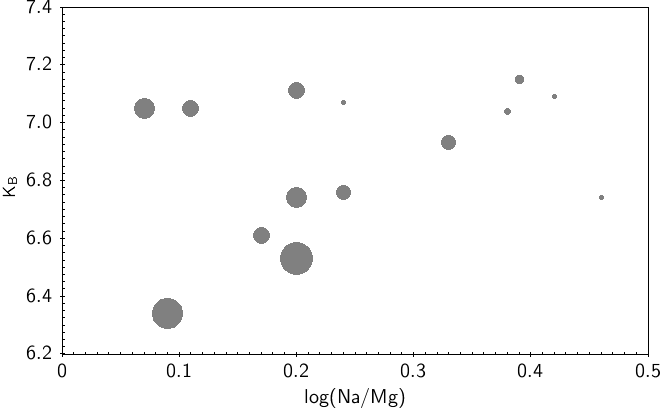}} \caption[f9]{The dependence of the Na/Mg line intensity ratio on material strength parameter $K_B$. The varying sizes of the marks display photometric masses of individual meteoroids in logarithmic scale.} \label{Kb_logNa_size}
\end{figure}

\subsection{Mineralogical density}
\label{}

Mineralogical density is certainly one of the most important physical properties defining the nature meteoroids. There are numerous methods for the determination of meteoroid densities. Bulk densities of meteoroids can be calculated from meteor observations by simultaneously fitting the deceleration and the light curves of meteors (e.g. Borovi\v{c}ka et al., 2007; Kikwaya et al., 2011). The effective fragmentation model must be taken into account.	

Mineralogical density represents the density of the substance, irrespective of the structure and shape of the body (Bronshten, 1983). The actual (mineralogical, or grain) density of the meteoroid material differs from its bulk (volume) density due to the porous structure, presence of voids and porosities, which are characteristic for most meteoroids. We applied the method suggested by Ceplecha (1958) and developed by Benyukh (1968), which uses the heat conductivity equation to obtain the mineralogical density of meteoroids. The equation of heat conductivity can be written in the form:
\begin{equation}
\frac{2T_B(\lambda \delta_m c)^{1/2}}{\Lambda} = \frac{v_\infty^{5/2} \rho}{(b \cos{z_r})^{1/2}},
\end{equation}
where $T_B$ is the temperature of the frontal surface of a meteoroid; $\lambda$ is the meteoroid heat conductivity; $\delta_m$ is the mineralogical density of the meteoroid; $c$ is the specific heat of the meteoroid; $\Lambda$ is the heat transfer coefficient in the beginning of evaporation; $v_\infty$ is the pre-atmospheric velocity of the meteoroid; $\rho$ is the atmospheric density; $z_R$ is the zenith distance of the radiant; and $b=1/H$ is the air density gradient, where $H$ is the scale height. The atmospheric parameters (densities, temperatures) were determined from the MSIS-E-90 Atmosphere Model of Hedin (1991). Values of parameters in Eq. 6 are applied at the meteor beginning height.

\begin{table}
\centering
\small\begin{center}\caption {Laboratory data of mineralogical density ($\delta_m$), specific heat ($c$), heat conductivity ($\lambda$), and corresponding function $f(\delta_m)$ (Eq. 7) for different materials (Berch et al., 1949).} \vspace{0.5cm}
\resizebox{\textwidth}{!}{\begin{tabular}{lcccccccccc}
\hline\\
&&&&&\\[-12pt]
\multicolumn{1}{l}{Material}& %
\multicolumn{1}{c}{$\delta_m$}& %
\multicolumn{1}{c}{$c$ x $10^7$}& %
\multicolumn{1}{c}{$\lambda$}& %
\multicolumn{1}{c}{$f(\delta_m)$}& \\
 & (g\,cm\textsuperscript{-3}) & (erg\,g\textsuperscript{-1}\,deg\textsuperscript{-1}) & (erg\,s\textsuperscript{-1}\, cm\textsuperscript{-1}\, deg\textsuperscript{-1}) &  \\
\hline\\
&&&&\\[-12pt]
Brick (Kizel'gur) & 0.40 &	0.75 & 9.3 x 10\textsuperscript{3} & 8.72  \\
Sand (river) & 1.52 &	0.70 & 2.7 x 10\textsuperscript{4} & 9.39 \\
Sandstone & 2.45 &	0.93 & 1.7 x 10\textsuperscript{5} & 9.93 \\
Schist & 2.67 & 0.71 & 1.5 x 10\textsuperscript{6} & 10.13 \\
Granite & 2.70 & 0.65 & 2.5 x 10\textsuperscript{5} & 10.15 \\
Basalt & 2.90 &	0.85 &	2.0 x 10\textsuperscript{5} & 10.20 \\
Fluorit & 3.20 & 0.85 & 1.0 x 10\textsuperscript{6} & 10.25 \\
Fyalit & 4.40 &	0.55 & 2.0 x 10\textsuperscript{6} & 10.52 \\
Gematit & 5.26 & 0.61 & 2.4 x 10\textsuperscript{5} & 10.66 \\
Arsenopirit & 6.07 & 0.43 &	3.8 x 10\textsuperscript{6} & 10.82 \\
Iron & 7.60 & 0.44 & 7.3 x 10\textsuperscript{6} & 11.00 \\
\hline
\end{tabular}}\label{Laboratory}
\end{center}
\end{table}

The laboratory data of $\lambda$, $\delta_m$, and $c$ are available for various types of rocks, minerals, and metals (Berch et al., 1949), applied values are given in Table 4. The values of $T_B$ and $\Lambda$ can be assumed for meteoric stone and iron particles. According to Levin (1956), the temperature of a meteoroid frontal surface at the beginning height equals 1600 K for friable stone particles, 2400 K for dense particles, and 2800 K for iron particles. The heat transfer coefficient $\Lambda$ is assumed to be 1 for stone meteoric particles and 0.75 for iron bodies (Levin, 1956). By applying these values and defining the decimal logarithm of the left-hand side of Eq. 6 as the function
\begin{equation}
f(\delta_m) = \log{\Big[\frac{2T_B(\lambda \delta_m c)^{1/2}}{\Lambda}\Big]},
\end{equation}
we are able to construct the dependence of $f(\delta_m)$ on $\delta_m$  (Figure 11). The decimal logarithm of the right-hand side of Eq. 6,
\begin{equation}
f(\delta_m) = \log{\Big[\frac{{v_\infty^{5/2}} \rho}{(b \cos{z_r})^{1/2}}\Big]},
\end{equation}
can be calculated from precise video observations and the applied atmosphere model. Using these values of $f(\delta_m )$ in the diagram on Figure 11, we are able to determine the mineralogical density $\delta_m$ of meteoroids.	

\begin{figure}
\centerline{\includegraphics[width=8cm,angle=0]{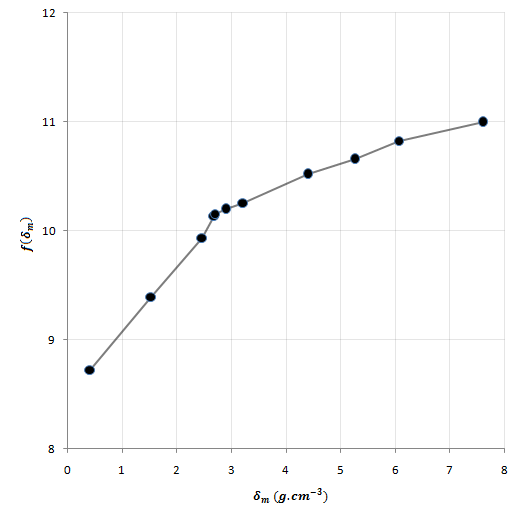}} \caption[f10]{The dependence of the function $f(\delta_m)$ ) (Eq. 7) on mineralogical density $\delta_m$. The points represent values for different materials assuming physical parameters, which were used by Benyukh (1968).} \label{min}
\end{figure}

This technique was used by Benyukh (1974) to estimate the mineralogical densities of over 3000 meteoroids from a database of meteors captured by Super-Schmidt cameras. Recently, Babadzhanov \& Kokhirova (2009) applied the aforementioned method to determine mineralogical densities of 501 meteoroids observed by photographic cameras, delivering results in satisfactory agreement with the results of Benyukh (1974). Considering the sensitivity of the AMOS system, which is similar to the sensitivity of Super-Schmidt cameras, we expect to observe comparable beginning heights of meteors and thus comparable results for mineralogical densities.

We determined high spread of mineralogical densities in our sample of Taurids ranging from as low as 1.3 g\,cm\textsuperscript{-3} to more expected values of approximately 2.5 g\,cm\textsuperscript{-3} (Table 5). While over half of the meteoroids are positioned close the upper limit of this scale, we also observe a non-negligible population of low density Taurids. It should be taken into account that the applied technique is highly dependent on the observed beginning height of meteors (Figure 12). Figure 12 reveals that the two most massive meteoroids in our sample are the ones with the lowest mineralogical density. Koten et al. (2004) shown that the beginning heights are functions of meteoroid mass and that this dependence can exhibit different behavior for different meteor showers. The works of Benyukh (1974) and Babadzhanov \& Kokhirova (2009) provided mineralogical densities for a sample of photographic Taurids with the mean value of 2.7 $\pm$ 0.2 g\,cm\textsuperscript{-3}. While the heterogeneity of the Taurid stream could provide such observed material variations, the results could also be affected by the slightly higher sensitivity of AMOS system, and the differences in the applied atmospheric model. Nevertheless, we identified compact population of Taurids positioned close to the modus of the mineralogical density values in our sample at around 2.5 g\,cm\textsuperscript{-3}. Meteoroids with higher mineralogical density expectedly also show higher values of material strength parameters $K_B$ and $P_E$. Whereas the relation between mineralogical density and $K_B$ is particularly apparent due to the common dependence on meteor beginning height, the $P_E$ - density functionality is more complex but still evident. 

\begin{figure}
\centerline{\includegraphics[width=8cm,angle=0]{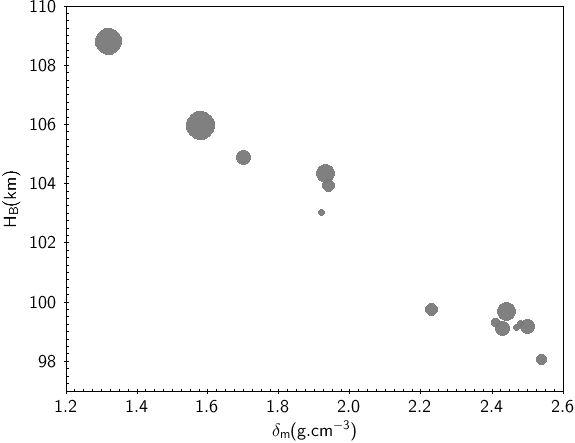}} \caption[f11]{The dependence between the mineralogical density and beginning height of Taurid meteoroids. The varying sizes of the marks display photometric masses of individual meteoroids in logarithmic scale.} \label{Min_dens_HB_logmp}
\end{figure}

\subsection{Dynamic pressure}
\label{}

The bulk strengths of meteoroids can be also inferred from the loading ram pressure causing their fragmentation. During the aerodynamic loading, the internal stresses in the meteoroid are proportional to the exerted ram pressure (Fadeenko, 1967), and so the stagnation pressure at breakup can be used as a measure of meteoroid strength in the disruption (Popova et al., 2011). The fragmentation strength of a meteoroid can be simply taken as the dynamic pressure acting at the front surface of the meteoroid:
\begin{equation}
p = \rho_h v_h^2,
\end{equation}
where $\rho _h$ and $v_h$ are the atmospheric density and meteor velocity at the height of the fragmentation. Both of these quantities can be easily measured from the observational data and the atmospheric model, but the determination of the fragmentation height is often difficult. Various methods can be applied for this purpose, depending on the type of available data and the type of meteoroid fragmentation (Popova et al., 2011).		

The fragmentation heights of the Taurid meteoroids presented here were determined from meteor light curves. The meteor flares (sudden transient increases of brightness) observed mainly in bright fireballs are caused by the sudden release of dust or small fragments from the meteoroid. The position of the flares in Taurid light curves was used to infer the fragmentation height. However, not all of the analyzed Taurid meteors exhibited recognizable flares, and therefore we were not able to determine reliable values of dynamic pressures for these, characteristically fainter cases. The inferred fragmentation strengths at first breakup of 8 Taurids for which we observed recognizable flares are in Table 5.	

Taurid fireballs were previously observed to break-up under 0.05 - 0.18 MPa (Konovalova, 2003). The inferred dynamic pressures in our sample of Taurid meteoroids ranges from 0.02 - 0.10 MPa (Table 5), with the average fragmentation strength of 0.05 MPa placed on the lower limit of the results of Konovalova (2003). Expectedly, meteoroids of higher mineralogical density fragmented under higher pressures than the low-density bodies. The majority of C type Taurids were characterized by low fragmentation strengths with very low values observed in the two D type meteors (0.024 and 0.034 MPa). While there are some discrepancies observed between the $K_B$ and $P_E$ classification with respect to determined dynamic pressures, the obtained values are in general characteristic for fragile cometary bodies. Tensile strengths of average small shower meteoroids were examined by Trigo-Rodr\'{i}guez \& Llorca (2006), who found that next to Geminids, the Taurids contained the strongest shower meteoroids, with bulk densities in some fireballs comparable to the Tagish Lake meteorite (Hildebrand et al., 2006). In a survey of instrumentally observed meteorite falls, Popova et al. (2011) found that these bodies show very low bulk strengths of 0.1 to 1 MPa on first breakup. The lower limits of this range are characteristic for the weakest observed meteorite falls - carbonaceous chondrites Tagish Lake, Maribo and Sutter's Mill (Borovi\v{c}ka et al., 2015). Based on the inferred dynamic pressures in our sample, we can assume that Taurid meteoroids are on average weaker compared carbonaceous chondrites. Still, the stream could include particles of comparable mechanical strengths, as the upper limits of the determined fragmentation pressures reach 0.1 MPa. The three instrumentally observed carbonaceous meteorite falls have shown several common characteristics. All objects were initially very large in size and showed features indicating very fragile structure (early fragmentation, high end heights). The very similar orbits of Maribo and Sutter's Mill suggest common and relatively recent origin (Jenniskens et al., 2012). The entry velocities of these meteorite falls are around 28 km/s, which is also the average speed of Taurid meteors, and very close to the cut-off velocity suggested by modeling for meteorite production (ReVelle, 1979). While the Taurids remain the most promising major meteor shower candidate for meteorite recovery (Brown et al., 2013; Madiedo et al., 2014; Brown et al., 2016), the low strengths of its meteoroids, though in the upper limits comparable to the carbonaceous chondrites, would require significant initial mass of the impacting body.

\begin{table}
\centering
\small\begin{center}\caption {Physical properties of the studied multi-station Taurid meteoroids. Each Taurid is designated with meteor number, absolute magnitude ($Mag$), beginning height ($H_B$), terminal height ($H_E$), Na/Mg intensity ratio, Fe/Mg intensity ratio, spectral class, material strength parameters ($K_B$, $P_E$) and corresponding material type, dynamic pressure ($p$) and mineralogical density ($\delta_m$). The uncertainties of photometeric masses are in the given order of magnitude; the accuracy of the beginning and terminal heights is on the order of the last digit. Meteor IDs are the same as in Table 1.} \vspace{0.5cm}
\resizebox{\textwidth}{!}{\begin{tabular}{crrrrcccrcrccc}
\hline\\
&&&&&&&&&&\\[-12pt]
\multicolumn{1}{c}{No.}& %
\multicolumn{1}{r}{$Mag$}& %
\multicolumn{1}{r}{$M_p(g)$}& %
\multicolumn{1}{r}{$H_B(km)$} & %
\multicolumn{1}{r}{$H_E(km)$} & %
\multicolumn{1}{c}{Na/Mg} & %
\multicolumn{1}{c}{Fe/Mg} & %
\multicolumn{1}{c}{Class}& %
\multicolumn{1}{r}{$K_B$}& %
\multicolumn{1}{c}{}& %
\multicolumn{1}{r}{$P_E$}& %
\multicolumn{1}{c}{}& %
\multicolumn{1}{c}{$p$(MPa)}& %
\multicolumn{1}{c}{$\delta_m(g\,cm^{-3})$} \\
\hline\\
&&&&&&&&&\\[-12pt]
1 & -4.7 &	40 & 104.0 & 70.1 & 1.74 & 0.65 &	Fe poor & 6.76 & C1 & -5.32 & IIIA & - & 1.94 \\
  & $\pm$ 0.7 &	 &  &  & 0.14 & 0.08 & & 0.04 & & 0.15 &  &  & 0.18 \\
2 & -4.5 &	60 & 99.2 & 52.7 & 1.59 &	1.18 & normal &	7.11 & B & -4.36 & I & - & 2.50 \\
  & $\pm$ 0.7 &	 &  &  & 0.16 & 0.13 & & 0.01 & & 0.13 &  &  & 0.05 \\
3 & -5.1 &	50 & 99.1 & 69.3 & 1.30 &	0.77 & normal &	7.05 & C1 &	-5.35 &	IIIA & 0.063 &	2.43 \\
  & $\pm$ 0.8 &	 &  &  & 0.14 & 0.10 & & 0.02 & & 0.15 &  & 0.004 & 0.15 \\
4 & -6.0 &  150 & 99.7 & 60.7 & 1.17 &	1.05 & normal &	7.05 & C1 &	-4.98 &	II & 0.099 & 2.44 \\
  & $\pm$ 1.2 &	 &  &  & 0.09 & 0.09 & & 0.01 & & 0.21 &  & 0.003 & 0.10 \\
5 & -8.1 &	1000 & 108.8 & 56.6 & 1.23 & 1.73 & normal & 6.34 & D & -5.13 & II & 0.034 & 1.32 \\
  & $\pm$ 2.0 &	 &  &  & 0.05 & 0.06 & & 0.02 & & 0.36 &  & 0.002 & 0.15 \\
6 & -2.1 &	1 &	99.3 &	72.2 &	2.61 & 1.70 & normal & 7.09 & C1 & -5.00 & II & - & 2.48 \\
  & $\pm$ 0.7 &	 &  &  & 0.50 & 0.37 & & 0.01 & & 0.11 &  &  & 0.07 \\
7 & -5.6 &	40	& - & - & 1.48 & 1.15 & normal & - & - & - & - & - & - \\
  & $\pm$ 2.0 &	 &  &  & 0.07 & 0.06 & & & & &  &  & \\
8 & -1.8 &	1 &	99.2 &	73.4 &	1.73 & 0.72 & normal & 7.07 & C1 & -5.04 & II & - & 2.47 \\
  & $\pm$ 0.7 &	 &  &  & 0.35 & 0.22 & & 0.02 & & 0.13 &  &  & 0.13 \\
9 & -8.4 &	1500 & 106.0 & 60.5 & 1.59 & 1.10 & normal & 6.53 & D & -5.51 & IIIA &	0.024 &	1.58 \\
  & $\pm$ 1.4 &	 &  &  & 0.07 & 0.06 & & 0.03 & & 0.35 &  & 0.002 & 0.15 \\
10 & -1.9 &	1 &	99.3 &	68.5 &	2.39 &	0.61 & Fe poor & 7.04 & C1 & -4.90 & II & - & 2.41 \\
  & $\pm$ 0.6 &	 &  &  & 0.22 & 0.10 & & 0.02 & & 0.11 &  &  & 0.13 \\
11 & -5.0 & 70 & 104.9 &	63.7 &	1.49 &	0.61 & Fe poor & 6.61 & C1 & -5.14 & II & 0.018 & 1.70 \\
  & $\pm$ 0.9 &	 &  &  & 0.06 & 0.04 & & 0.02 & & 0.16 &  & 0.002 & 0.08 \\
12 & -5.5 & 150 & 104.4 & 68.1 & 1.57 & 0.80 &	normal & 6.74 &	C1 & -5.17 & II & 0.027 & 1.93 \\
  & $\pm$ 0.6 &	 &  &  & 0.09 & 0.06 & & 0.01 & & 0.11 &  & 0.002 & 0.13 \\
13 & -1.3 &	1 &	103.0 & 62.8 & 2.90 &	0.60 & Na enhanced & 6.74 &	C1 & -4.46 & I & - & 1.92 \\
  & $\pm$ 0.9 &	 &  &  & 0.38 & 0.16 & & 0.02 & & 0.16 &  &  & 0.08 \\
14 & -3.9 &	50 & 99.8 & 54.5 & 2.12 & 0.64 & Fe poor & 6.93 &	C1 & -4.42 & I & 0.054 & 2.23 \\
  & $\pm$ 0.9 &	 &  &  & 0.09 & 0.05 & & 0.02 & & 0.14 &  & 0.003 & 0.02\\
16 & -2.9 &	10 & 98.1 &	57.0 &	2.44 & 0.39 & Na enhanced &	7.15 & B & -4.49 & I & 0.081 & 2.54 \\
  & $\pm$ 0.7 &	 &  &  & 0.23 & 0.09 & & 0.01 & & 0.11 &  & 0.002 & 0.03\\

\hline
\end{tabular}}\label{Physical}
\end{center}
\end{table}

\section{Conclusion}
\label{}

We presented an analysis of spectral and physical properties of Taurid meteoroids observed during the outburst in November 2015. In a sample of 14 multi-station Taurid meteors with spectra, we identified 10 Southern Taurids and 4 meteors belonging to the Northern Taurids shower. The Tisserand's parameter defined the orbital origin of the sample on the borderline between asteroidal orbits (10 meteors) and Jupiter-family type cometary orbits (4 meteors), as would be expected for a stream originating in the short-period comet 2P/Encke. Furthermore, we used a simple method based on the Southworth-Hawkins $D_{SH}$ orbital similarity criterion to look for associations with related Taurid complex NEOs. We found possible association to 10 asteroids satisfying the similarity threshold value of $D_{SH} \leq$ 0.10. The most meteor associations in the southern branch of our sample was found with asteroids 2015 TX24, 2003 UV11, and 2007 UL12. Meteoroids from the northern branch showed most markable similarities with asteroid 2014 NK52. However, we argue that based on the determined spectral and physical properties of Taurid meteoroids, these associations may be purely incidental. 

The spectral analysis revealed large dispersion of iron content in Taurid meteoroids, ranging from Fe-poor bodies to meteoroids with almost chondritic Fe/Mg ratios. Na content also varied slightly and was found to be a function of the meteoroid perihelion distance, assigning Na loss to space weathering processes, particularly the thermal desorption. It was also found that the Na enrichment is preferred in smaller particles. Nevertheless, the increased Na line intensity over Mg was found in all of the studied spectra, indicating significant volatile content in Taurid meteoroids. The overall spectral classification of Taurid meteors observed in Slovakia and Chile showed corresponding results dominated by normal-type spectral classes and several cases of Fe-poor and Na-enhanced meteoroids. Besides the almost chondritic content of iron in few of the studied cases, there are no spectral features clearly suggesting asteroidal origin in any of the analyzed meteoroids.

Overall, the determined material strengths showed typically cometary characteristics in the $K_B$ classification, dominated by C group meteoroids, and much more dispersed values in the $P_E$ classification, in which the majority of Taurid meteoroids fall in the group II, characteristic for carbonaceous bodies. The discovered features suggest that Taurids are heterogeneous population of meteoroids, which are cometary in nature but contain solid, possibly carbonaceous inclusions. The diversity of the observed material may be caused by the reprocessing of meteoroid surfaces, but possibly also reflects the authentic inhomogeneity of 2P/Encke or earlier parent comet of the stream. Four meteoroids of group I characterized with higher mineralogical densities were identified as the best possible candidates for non-cometary origin, however, the asteroidal source cannot be confirmed in any of the meteoroids in our sample. Moreover, we also found high dispersion of mineralogical densities among the Taurid meteoroids. Obtained very low densities in individual cases were probably influenced by the high beginning heights of these meteors, which relate to the sensitivity of the observational system. Nevertheless, the largest population was identified with mineralogical densities around 2.5 g\,cm\textsuperscript{-3}, which is comparable to the mean values of 2.7 g\,cm\textsuperscript{-3} determined by Benyukh (1974) and Babadzhanov \& Kokhirova (2009). Mechanical strength of several Taurids, which exhibited bright flares, was also estimated from the dynamic pressure causing their fragmentation. It was found that studied meteoroids started fragmenting under low pressures of 0.02 - 0.10 MPa. While these values suggest that Taurid meteoroids are on average weaker compared to the instrumentally observed carbonaceous meteorite falls (first breakup at 0.1 - 1 MPa), the stream could include bodies of strength comparable to carbonaceous chondrites, as the upper limits of dynamic pressures in our sample reach 0.1 MPa.


\hspace {0.5cm}\hspace {0.5cm}

\textbf{Acknowledgements.} This work was supported by the Slovak Research and
Development Agency under the contract No. APVV-0516-10, No. APVV-0517-12, and by the Slovak Grant Agency
for Science, grants No. VEGA 1/0225/14.


\hspace {0.5cm}\hspace {0.5cm}

\bibliographystyle{elsarticle-harv}
\bibliography{<your bibdatabase>}


\end{document}